\documentclass[10pt]{article}

\usepackage[preprint]{tmlr}

\newif\ifnamedversion
\newif\ifpreprintversion
\makeatletter
\if@accepted\namedversiontrue\else\namedversionfalse\fi
\if@preprint\preprintversiontrue\else\preprintversionfalse\fi
\makeatother

\usepackage{amsmath,amssymb}
\usepackage{graphicx}
\usepackage{booktabs}
\usepackage{multirow}
\usepackage{xcolor}
\usepackage{hyperref}
\ifnamedversion
  \ifpreprintversion
    \hypersetup{colorlinks=true,linkcolor=blue!50!black,citecolor=blue!50!black,urlcolor=blue!50!black,
      pdftitle={FlyHack: Stress-Testing Mechanistic Claims in a Connectome-Constrained Visual Model},
      pdfauthor={Aashish Dhawan},pdfsubject={Preprint}}
  \else
    \hypersetup{colorlinks=true,linkcolor=blue!50!black,citecolor=blue!50!black,urlcolor=blue!50!black,
      pdftitle={FlyHack: Stress-Testing Mechanistic Claims in a Connectome-Constrained Visual Model},
      pdfauthor={Aashish Dhawan},pdfsubject={TMLR camera-ready article}}
  \fi
\else
  \hypersetup{colorlinks=true,linkcolor=blue!50!black,citecolor=blue!50!black,urlcolor=blue!50!black,
    pdftitle={FlyHack: Stress-Testing Mechanistic Claims in a Connectome-Constrained Visual Model},
    pdfauthor={Anonymous Authors},pdfsubject={Anonymous TMLR submission}}
\fi
\usepackage{url}
\usepackage[capitalise,noabbrev]{cleveref}

\newcommand{\Fc}{F}
\newcommand{\TF}{F_{\mathrm{tun}}}

\title{FlyHack: Stress-Testing Mechanistic Claims in a\\ Connectome-Constrained Visual Model}

\ifnamedversion
\author{\name Aashish Dhawan \\
      \addr Department of Computer \& Information Science \& Engineering\\
      University of Florida}
\else
\author{\name Anonymous Authors}
\fi

\def\month{MM}
\def\year{YYYY}
\def\openreview{\url{https://openreview.net/forum?id=XXXX}}

\begin{document}
\maketitle
\begingroup
\renewcommand{\thefootnote}{\fnsymbol{footnote}}
\footnotetext[1]{AI assistants were used for software development and manuscript editing. The authors determined the research questions, experimental design, analyses and scientific claims; verified all reported results and references; and take full responsibility for the work.}
\endgroup

\begin{abstract}
Connectome-constrained neural networks make biological components directly accessible to intervention, but their mechanistic conclusions can still depend on the comparison model, intervention, circuit representation and trained solution. We stress-test these dependencies in flyvis, a \emph{Drosophila} motion-vision model, using its 50-model ensemble, task-matched causal CNN controls and replicated interventions. At the level of model behaviour, three-level reverse-phi reversal in flyvis is strongly period- and velocity-dependent and also appears in task-matched CNNs; neither family reproduces the fly's stationary-sawtooth response. Adversarial direction reversal requires a perturbation bound of 0.21--0.27 times stimulus contrast in flyvis versus 0.13--0.25 in CNNs, but the gap over a parameter-matched CNN largely closes for larger CNNs. At the level of mechanism, matching attribution to the intervention matters: for held-out-mean replacement, first-order edge attribution patching reaches pooled AUROC 0.970 and Spearman 0.872, versus 0.933 and 0.126 when its zero-aligned score is evaluated against the same ground truth. Circuit conclusions also vary with representation and ensemble member. Greedy cell-type sets meet a model-relative T4/T5 tuning criterion in 25 of 50 models, while T4/T5 output silencing reduces the tested optic-flow behaviour to at most two of eight clips in 40. In six models where the cell-type search first succeeds, the corresponding connection sets can be pruned to 13--31 added classes; under an additional two-sided response-amplitude check, only two of those six first-success cell-type sets still meet the tuning threshold. These results support a narrower use of connectome-constrained models: they make mechanistic hypotheses directly testable, but the resulting claims must be scoped to the model family, intervention, representation and trained solution.
\end{abstract}

% ===========================================================================
\section{Introduction}
\label{sec:intro}

Recent connectomes now span the adult \emph{Drosophila} brain, visual system and male central nervous system \citep{dorkenwald2024,nern2025,berg2026}, making it possible to build network models whose architecture is constrained by anatomy and whose remaining parameters are fitted by simulation or training \citep{lappalainen2024,shiu2024}. The most developed example in vision is flyvis \citep{lappalainen2024}: a recurrent network of 64 optic-lobe cell types, with connectivity constrained by the connectome, whose 734 free parameters were trained to estimate optical flow. Its trained ensemble predicts physiological tuning in many cell types without having been fitted to neural recordings. Such models are increasingly used to generate in-silico predictions about circuit function and behaviour.

The central question in this paper is not whether flyvis is biologically constrained---it is---but how much a mechanistic conclusion changes when we change the comparison model, intervention, circuit representation or trained ensemble member. We address that question in two stages.

First, we test model-family behaviour. We train causal CNNs on the same task, Sintel rendering and target, loss and sampled input lattice at three sizes, from a parameter-matched 8{,}161 parameters to about one million, and compare motion illusions and adversarial sensitivity. These are controls at the level of model families, not topology-only ablations: recurrence, learned time constants, type-shared parameters, readout and training schedule remain confounded with connectome wiring. The CNN training split is documented, whereas the released flyvis models' launch-time split is not recoverable; their common evaluation is therefore descriptive rather than a shared held-out generalization comparison.

Second, we turn from behaviour to mechanism within flyvis. Each of its 604 type-to-type connection classes and each modelled cell type can be intervened on directly. This lets us score circuit-discovery outputs against exhaustive single-component interventional ground truth rather than against a planted circuit \citep{lindner2023,gupta2024} or a faithfulness proxy \citep{mueller2025}, and then ask whether the answer changes with intervention semantics, circuit representation or ensemble member. Directional predictions and analysis criteria were specified before the corresponding analyses; they were not externally timestamped preregistrations, and we report failed predictions as well as successful ones (\cref{app:prereg}).

\paragraph{Contributions.}
\begin{enumerate}
\item \textbf{Model-family behaviour (\cref{sec:illusions,sec:attacks}).} Under a \emph{Drosophila}-matched three-level reverse-phi stimulus, flyvis and task-matched CNNs both show period- and velocity-dependent reversal; neither family reproduces the fly's stationary-sawtooth response. flyvis has a larger adversarial margin than a parameter-matched CNN, but the gap largely closes for larger CNNs.
\item \textbf{Attribution against exhaustive interventions (\cref{sec:circuits}).} Same-stimulus mean replacement leaks stimulus information, so held-out mean replacement is primary. Matching integrated gradients and edge attribution patching to the intervention path substantially improves continuous-effect agreement, especially for the latter.
\item \textbf{Circuit representation and ensemble dependence (\cref{sec:types}).} Under one T4/T5 tuning criterion, cell-type solutions with 4--9 added types correspond to connection sets that can be pruned to 13--31 added classes (different intervention granularities, not an equal-unit comparison); across the 50-model ensemble these solutions and the effect of T4/T5 silencing differ between models.
\item \textbf{Evaluation artefacts (\cref{app:artefacts}).} Two-level periodic reverse-phi constructions are phase-equivalent to translation, a one-sided retained-output metric is satisfied by degenerate sub-networks, and same-input mean-current ablation leaks stimulus information.
\end{enumerate}

% ===========================================================================
\section{Related work}
\label{sec:related}

\paragraph{Connectome-constrained models.}
\citet{lappalainen2024} trained ensembles of connectome-constrained networks of the fly visual system on optic flow and compared predicted tuning to published recordings; \citet{shiu2024} simulated the central brain as a leaky integrate-and-fire network with connectome weights. Anatomically constrained task optimization can also recover response properties of individual fly motion neurons \citep{mano2021}. \citet{karaneen2026} retrained flyvis ensembles and found that similar task performance does not reliably identify the circuit solutions whose tuning matches experiment. \citet{beiran2025} showed that connectivity alone often leaves the dynamics of connectome-constrained networks underdetermined, and that recordings from a small, well-chosen set of neurons can resolve this. Non-identifiability of circuit explanations also arises in fully enumerable networks \citep{meloux2025}, and degenerate circuit parameters have long been documented in small biological circuits \citep{prinz2004,marder2006}. Our analyses are complementary to the tuning comparisons of \citet{karaneen2026}: we intervene on the released ensemble and score attribution and circuit-search outputs against those interventions.

\paragraph{Motion illusions.}
Reverse-phi, the reversal of perceived direction when contrast inverts at each displacement \citep{anstis1970}, is predicted by correlation-type detectors \citep{hassenstein1956} and has been characterized in fly behaviour and T4/T5 physiology \citep{tuthill2011,leonhardt2017}. Minimal apparent-motion experiments show complementary bright/dark computations and ON/OFF-pathway contributions to phi and reverse-phi responses \citep{clark2011,salazargatzimas2018}. Flies turn in response to stationary sawtooth luminance gradients, an effect that depends on T4 and T5 \citep{agrochao2020}; humans perceive motion in related static patterns such as the peripheral-drift and Rotating Snakes illusions \citep{conway2005}. \citet{rosario2026} note that recent DNN motion models can reproduce dynamic illusions such as reverse-phi, but found that most tested optical-flow networks fail to produce human-consistent flow on Rotating Snakes. \citet{golan2020} argue that stimuli which separate candidate models are more informative than stimuli on which they agree.

\paragraph{Robustness and biological priors.}
Small input perturbations can change the outputs of deep networks \citep{szegedy2014,madry2018}, and constraining the early stages of a CNN to resemble primate V1 increases robustness \citep{dapello2020}. We compare the robustness of flyvis with task-matched CNN families and treat any difference as a model-family effect.

\paragraph{Circuit discovery and its evaluation.}
Circuit discovery identifies components responsible for a behaviour by activation patching \citep{wang2023ioi,zhang2024}, automated pruning \citep{conmy2023}, or gradient-based approximations such as attribution patching and integrated gradients \citep{syed2024,hanna2024,sundararajan2017}. Evaluation relies on planted circuits \citep{lindner2023,gupta2024} or on faithfulness metrics \citep{mueller2025}; \citet{miller2024} showed that faithfulness scores depend on the ablation type, and gradient-based attribution patching has been evaluated against exhaustive patching in language models \citep{kramar2024}. We evaluate these methods in a trained recurrent network in which single-component interventional ground truth can be computed exhaustively at both the connection-class and the cell-type level.

% ===========================================================================
\section{Model, controls and protocol}
\label{sec:setup}

\paragraph{flyvis.}
We use the 50 pretrained models released by \citet{lappalainen2024} (flyvis 1.2.0). The model represents 64 biological cell types across 721 hexagonal columns (about 45{,}000 neurons and 1.5 million neuron-level connections). CT1 is represented by separate medulla and lobula compartments, giving 65 parameterized type units. The 734 free network parameters comprise 65 time constants, 65 resting potentials and 604 synapse-strength scaling factors, one per type-to-type connection class; each class is a scale factor shared across all neuron-level contacts of that type pair, not an individual synapse. We call these 604 classes \emph{connections}. A trained decoder (DecoderGAVP, 7{,}427 parameters) reads 34 type units and outputs optical flow on the hexagonal lattice. The models were trained on Sintel \citep{butler2012}, and stimuli are applied through flyvis's own renderer. Experiments that are too expensive for the full ensemble use models 000--009, the ten models ranked best by the release package's legacy task-error scalar. That scalar is numerically EPE-like, but its exact generator, configuration, checkpoint-selection history and training-held-out status are not retained (\cref{app:validation}).

\paragraph{Readout.}
The readout is the decoded flow vector at the central column, averaged over an analysis window that begins after the onset transient and spans whole stimulus cycles (at least 0.3\,s). The decoded direction is the angle of this vector.

\paragraph{CNN controls.}
The CNNs receive the same luminance input on the 721-column lattice, mapped to a zero-padded $31\times31$ grid by a permutation of axial coordinates, so a per-column perturbation bound means the same thing for both families. For the common evaluation in \cref{tab:models}, flyvis and the CNNs use the original 16-subsequence Sintel evaluation set, preprocessing, optic-flow target, $\ell_2$-norm metric and batch size, with augmentation disabled. The CNNs were trained on the corresponding scene-disjoint original training split. The released flyvis package does not retain enough launch provenance to establish its training split; the best-supported random-fold reconstruction overlaps the original evaluation in 11 of 16 subsequences and all six scenes. The CNNs are seven-layer causal temporal convolutional models with a 15-frame (300\,ms) receptive field and direct flow heads, at three sizes (8{,}161 parameters, matched to flyvis network plus decoder; 50{,}002; 988{,}418) with five seeds each, trained with Adam for 30{,}000 steps (\cref{app:details}). flyvis checkpoints were trained for 250{,}000 steps. The controls do not isolate wiring: flyvis also differs in recurrent dynamics, learned time constants, type-level parameter sharing, decoder, effective temporal context and optimization schedule. Below, ``matched'' refers to task, Sintel rendering and target, loss, input lattice and, for the small CNN, parameter count; it does not mean a verified shared training split (\cref{tab:models}).

\begin{table}[t]
\centering
\caption{Models used in the attack and detailed circuit experiments. Common-evaluation $\ell_2$ norm is the equal-weight mean over the original 16-subsequence Sintel evaluation set with augmentation disabled; full preprocessing and reduction details are in \cref{app:validation}. The set is scene-held-out relative to CNN training but is not demonstrably held out for released flyvis, so this is a descriptive cross-family comparison rather than a fair shared held-out generalization test. The zero-flow predictor is 1209.6. Clean passes count the five grating contrasts (1 to 0.05) decoded within $45^\circ$ of truth.}
\label{tab:models}
\small
\begin{tabular}{lrrrr}
\toprule
model & parameters & $n$ & common-eval. $\ell_2$, mean (range) & clean passes \\
\midrule
flyvis 000--009 & 734 + 7{,}427 & 10 & 1132.7 (1103.5--1159.8) & 5 each \\
CNN small  & 8{,}161   & 5 & 1098.7 (1078.4--1114.7) & 5, 4, 5, 3, 4 \\
CNN medium & 50{,}002  & 5 & 1040.6 (1023.9--1053.9) & 5 each \\
CNN large  & 988{,}418 & 5 & 1028.8 (1013.0--1045.7) & 5 each \\
\bottomrule
\end{tabular}
\end{table}

\paragraph{Interventions.}
\emph{Connection ablation} sets a type-to-type synapse strength to zero. We distinguish four \emph{intervention semantics} by what replaces the removed current: nothing (zero); the connection's time-averaged current on the same target stimulus (same-stimulus mean, used only as a leakage diagnostic); a held-out mean computed from six calibration movies at $90^\circ$, $180^\circ$ and $270^\circ$, never the $0^\circ$ target movies; and the frame-by-frame current from the other target behaviour (counterfactual). Single-connection interventions start from a common fade-in state computed with the intact network. \emph{Cell-type silencing} multiplies a type's output, into both its outgoing currents and the decoder, by a gain inside the solver; the wide-field amacrine Am is always left intact. In the matched comparison of \cref{sec:types}, cell-type gains and connection masks are both applied during fade-in and stimulus.

\paragraph{Numerical reliability.}
Every intervention was evaluated three times; consensus effects take the median clean and intervened behaviour components separately and then normalize. Across 80 model--behaviour--semantic groups, every replicate pair preserved the same top-20 and top-40 connection sets, and the minimum full-vector replicate Spearman was 0.9823. We therefore interpret high-effect sets and threshold verdicts, not the ordering of near-zero effects.

% ===========================================================================
\section{Illusions}
\label{sec:illusions}

\subsection{Reverse-phi}

Reverse-phi reverses apparent direction when contrast polarity changes across displacements, and fly experiments implicate interactions between bright- and dark-motion pathways \citep{tuthill2011,leonhardt2017,clark2011,salazargatzimas2018}. Our stimulus follows \citet{leonhardt2017}. Alternating grating stripes stay at intermediate gray, while the other stripes are dark or bright. In matched standard trials the non-gray stripes keep their starting polarity; in reverse-phi trials they switch polarity at every $3.944^\circ$ spatial update. We test both starting polarities and average their response vectors before decoding. Spatial periods are $30^\circ$ and $90^\circ$, and update intervals of 1, 2, 3, 5, 10 and 25 frames at 50\,Hz give 197.2, 98.6, 65.7, 39.4, 19.7 and 7.9$^\circ$/s. Matched controls differ only in the per-update polarity switch.

\begin{figure}[t]
\centering
\includegraphics[width=\linewidth]{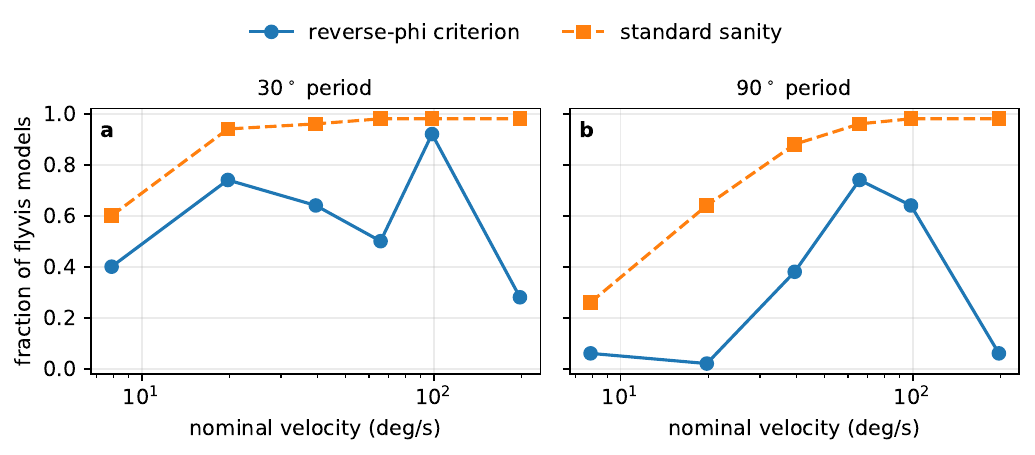}
\caption{Three-level reverse-phi evaluation with a fixed $3.944^\circ$ displacement per update. Solid: fraction of flyvis models meeting the reversal criterion; dashed: fraction passing the matched standard control (legend: ``standard sanity'').}
\label{fig:revphi}
\end{figure}

This three-level stimulus is not a translated copy of its matched control: none of 836 polarity-flipped frames equals any integer translation of the control frame. The residual is, however, a spatially uniform luminance offset. After subtracting each frame's spatial mean, the moving pattern equals a two-level contrast-inverting grating at half contrast, and a two-level contrast inversion is itself equivalent to a half-period translation (\cref{app:artefacts}). The three-level stimulus therefore adds a whole-field flicker and makes bright-only and dark-only views signal the nominal direction, but it does not remove the reverse-direction pattern correspondence that any motion-energy estimator can use. A model counts as reversal-positive when its standard control decodes within $45^\circ$ of nominal in at least three of four directions and its reverse-phi movie decodes more than $90^\circ$ from nominal in at least three of four directions.

flyvis reversal is structured rather than uniform (\cref{fig:revphi}). At $30^\circ$, counts across descending velocities are 14, 46, 25, 32, 37 and 20 of 50; at $90^\circ$ they are 3, 32, 37, 19, 1 and 3. The standard control passes in 47--49 models at $30^\circ$ except at the slowest velocity (30), and in 49, 49, 48, 44, 32 and 13 at $90^\circ$, which limits interpretation of the slowest conditions. Task-matched CNNs also reverse in selected bands: all five small and all five medium checkpoints reverse at $90^\circ$ and 65.7$^\circ$/s, whereas no CNN checkpoint reverses at $90^\circ$ and 197.2$^\circ$/s (\cref{tab:revphi_oldnew}). A reverse-phi-like response is therefore present under the fly-matched stimulus, but it is neither uniform nor unique to the connectome-constrained family, and it does not establish that the models implement the fly's mechanism. At $30^\circ$, the drop from 46/50 at $98.6^\circ$/s to 25/50 at $65.7^\circ$/s is not a control failure (49/50 pass at both velocities) and remains unexplained.

\subsection{Stationary sawtooth gradients}

Tethered flies turn in response to stationary sawtooth luminance gradients with a $45^\circ$ spatial period presented for 5\,s, towards the lighter shade along the shallow gradient, and the response is strongly reduced when T4 and T5 are silenced \citep{agrochao2020}. We rendered this stimulus in both polarities after 1\,s of mean grey, with normalized luminance in $[0,1]$ because the projector luminance values were not available. The prediction recorded before the run was that at least 35 of 50 flyvis models report the fly's turning direction in the sustained window, with magnitude at least 10\% of the response to a moving grating of matched contrast, and that the sign flips with polarity. It failed. Three of 50 flyvis models meet the criterion for each polarity and none for both; median sustained responses are 2.3--2.6\% of the moving-grating response, and no CNN meets the criterion (\cref{tab:static}). In 36/50 flyvis models the horizontal-flow sign is the same for the two polarities, and the polarity-flipping component has the fly-like sign in 23/50, close to the 25/50 expected by chance. With the four-level luminance sequence of the human peripheral-drift illusion \citep{conway2005}, at most 14 of 50 flyvis models and 2 of 6 CNNs report the predicted direction. This negative result is consistent with the optic-flow objective suppressing a static-gradient imbalance, but our experiments do not identify that objective as the cause. Together, the illusion probes do not provide a family-specific signature, so we next ask whether the families differ in sensitivity to input perturbations.

% ===========================================================================
\section{Adversarial perturbations}
\label{sec:attacks}

\paragraph{Attack and controls.}
Let $x$ be the clean input movie and $\mathbf{u}(x)$ the decoded flow at the central column, averaged over the analysis window. For a stimulus moving in direction $\hat{\mathbf{d}}$ we minimise
\begin{equation}
\mathcal{L}(\delta) = \cos\!\big(\mathbf{u}(x+\delta),\,\hat{\mathbf{d}}\big)
\quad\text{subject to}\quad \|\delta\|_\infty\le\epsilon,\; 0\le x+\delta\le 1,
\end{equation}
using projected gradient descent with a random start, step size $\epsilon/10$ and 60 steps (the loss plateaued within 20 steps in a pilot). The attack succeeds if the decoded direction is more than $90^\circ$ from $\hat{\mathbf{d}}$. The minimal successful bound $\epsilon^\ast$ is found by bisection in log space over $[0.002,0.5]$ to 10\% precision, so reported values lie on a coarse common grid. Three controls accompany each attack: uniform noise of amplitude $\epsilon^\ast$ (10 draws per cell); a \emph{naive} counter-moving grating of the same period and frequency, whose minimal amplitude $\epsilon_{\text{naive}}$ gives the attack advantage $\epsilon_{\text{naive}}/\epsilon^\ast$; and transfer of each successful perturbation to other models. Stimuli are gratings at 4\,Hz and contrasts 1, 0.5, 0.25, 0.1 and 0.05. Cells in which the clean stimulus is decoded incorrectly are excluded.

\begin{figure}[t]
\centering
\includegraphics[width=\linewidth]{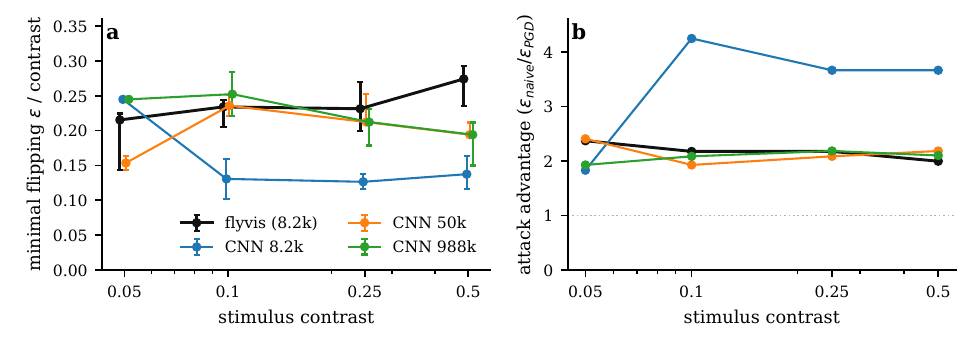}
\caption{Adversarial margins. (a) Minimal reversing $\ell_\infty$ perturbation divided by contrast (median and interquartile range). (b) Attack advantage over a naive counter-moving grating; the panel labels $\epsilon^\ast$ as $\epsilon_{\text{PGD}}$.}
\label{fig:margins}
\end{figure}

\paragraph{Margins.}
\Cref{tab:margins} and \cref{fig:margins} report $\epsilon^\ast$ divided by contrast. For flyvis this ratio lies between 0.21 and 0.27 at contrasts 0.05 to 0.5. For the parameter-matched CNN it is 0.14 and 0.13 at contrasts 0.5 and 0.25 (five seeds), so flyvis requires about twice the perturbation. The medium and large CNNs require 0.15--0.25, within a factor of 1.4 of flyvis, and the large CNN's median exceeds flyvis's at the two lowest contrasts (two seeds). At contrast 1, no bound up to 0.5 reverses any of the ten flyvis models, whereas three of the six CNN checkpoints tested at that contrast are reversed ($\epsilon^\ast=0.25$, 0.30 and 0.50, one per size). Uniform noise at $\epsilon^\ast$ reversed none of 660 draws. The attack advantage is close to 2 for flyvis and the larger CNNs, and 3.7--4.3 for the small CNN at contrasts 0.1 to 0.5. A square-wave grating of contrast $c$ deviates from mean luminance by $\pm c/2$, so a counter-moving component of amplitude near $c/2$ cancels it; the roughly constant $\epsilon^\ast/c$ and an advantage near 2 are consistent with the optimized perturbation concentrating opposite-direction motion energy at sensitive stimulus features.

\begin{table}[t]
\centering
\caption{Minimal reversing perturbation as a fraction of contrast (median over models and seeds; $n$ in parentheses) and attack advantage $\epsilon_{\text{naive}}/\epsilon^\ast$ (median, first two CNN seeds). Contrasts 0.5 and 0.25 pool five CNN seeds; Contrasts 0.1 and 0.05 use seeds 0 and 1 of each size, subject to the same clean-decoding exclusion. Identical entries arise because $\epsilon^\ast$ lies on the bisection grid.}
\label{tab:margins}
\small
\begin{tabular}{lcccc|cccc}
\toprule
 & \multicolumn{4}{c|}{$\epsilon^\ast/c$} & \multicolumn{4}{c}{attack advantage} \\
contrast $c$ & 0.5 & 0.25 & 0.1 & 0.05 & 0.5 & 0.25 & 0.1 & 0.05 \\
\midrule
flyvis     & 0.274 (10) & 0.231 (10) & 0.234 (10) & 0.215 (10) & 1.99 & 2.17 & 2.17 & 2.37 \\
CNN small  & 0.137 (5)  & 0.126 (5)  & 0.131 (2)  & 0.245 (1)  & 3.66 & 3.66 & 4.25 & 1.83 \\
CNN medium & 0.194 (5)  & 0.212 (5)  & 0.236 (2)  & 0.154 (2)  & 2.18 & 2.08 & 1.93 & 2.40 \\
CNN large  & 0.194 (5)  & 0.212 (5)  & 0.252 (2)  & 0.245 (2)  & 2.10 & 2.18 & 2.08 & 1.93 \\
\bottomrule
\end{tabular}
\end{table}

\paragraph{Transfer.}
Over all 50 flyvis models and six CNNs, transfer rates show no consistent family ordering: perturbations scaled to each source's own $\epsilon^\ast$ transfer less often between flyvis models (0.15) than between same-size CNN seeds (0.29), while at a fixed bound the order reverses (0.61 versus 0.50). Input gradients are more similar among flyvis models (mean cosine 0.14) than among same-size CNN seeds ($-0.08$), but this similarity does not produce a consistent transfer advantage (\cref{app:transfer}). These family-level tests therefore motivate a different use of flyvis's biological structure: rather than treating behaviour as evidence for wiring, we use the model's named components to test mechanistic explanations directly.

% ===========================================================================
\section{Circuit discovery with an exhaustive single-component ground truth}
\label{sec:circuits}

\subsection{Behaviours and replicated ground truths}

We study two behaviours on models 000--009. B0 is the decoded flow component along the true direction for a 4\,Hz standard grating. B1 is the negative decoded component along the nominal displacement direction for a two-level contrast-inverting grating ($30^\circ$ period, $7.9^\circ$/s; \cref{app:artefacts}), which is positive in all ten models; it is a second, direction-reversed readout rather than the three-level stimulus of \cref{sec:illusions}. The clean B1 component is small relative to B0 in several models: B1 ranges from 0.0147 to 2.007 while B0 ranges from 3.406 to 25.503, and in six models B1 is about 2.1\% of B0 or less, so normalization by $M_{\text{clean}}$ is ill-conditioned for some B1 effects. For each of the 604 connections and each intervention semantic, the effect is
\begin{equation}
e_j = \frac{M_{\text{clean}} - M_{\setminus j}}{M_{\text{clean}}}.
\end{equation}

Same-stimulus mean replacement is not an independent ground truth because it re-injects stimulus-specific information (\cref{app:artefacts}), so \emph{held-out mean} replacement is primary. The difference is moderate for the high-effect sets: per model, the two share on average 16.1 of the top 20 connections for B0 (range 14--18) and 15.5 for B1 (range 12--18). Continuous effect magnitudes differ more, with mean per-model Spearman 0.785 for B0 and 0.699 for B1. We call the intersection of the top-20 sets under zero, mean and counterfactual replacement the \emph{stable core} of a model--behaviour pair (a ranking-stability construct, not a uniquely identified mechanism); its mean size changes little between same-stimulus and held-out means (6.0 to 6.6 connections for B0; 5.6 to 5.3 for B1). The inhibitory fraction of ground-truth top-20 sets falls from 0.480 to 0.460 for B0 and from 0.485 to 0.430 for B1, against 0.377 over all connections.

\subsection{Intervention-matched attribution}

We evaluate integrated gradients \citep[IG;][]{sundararajan2017} and first-order edge attribution patching \citep[EAP;][]{syed2024}. Because component importance depends on the ablation or replacement baseline \citep{li2024optimal,chan2022causal}, we treat zero, held-out-mean and counterfactual replacement as distinct causal questions. The zero-aligned IG path varies all synapse strengths from zero to their trained values, and first-order EAP linearizes the effect of zeroing an edge. For held-out and counterfactual semantics, matched IG uses a \emph{global semantic path}: all 604 strengths are scaled together as $w(\alpha)=\alpha w$ while the summed replacement-current field is removed as $x(\alpha)=x+(1-\alpha)R$, where $x$ is the per-node input to the recurrent dynamics and $R=\sum_j r_j$ sums the replacement currents over connections. We approximate the path integral with 20 midpoint steps and assign each connection its weight-path and replacement-input contributions. Matched first-order EAP uses the clean-endpoint approximation $\nabla_w M\,w-\langle\nabla_x M,r_j\rangle$. The matched IG path aligns the global baseline with the intervention family but is not identical to integrating 604 separate one-connection paths; the ground truth remains one-connection replacement. Attributions are computed three times and aggregated by the median; across 120 model--behaviour--method--semantic groups the minimum replicate Spearman is 0.999976 and top-20 membership is identical.

\begin{table}[t]
\centering
\caption{Attribution against exhaustive single-connection ground truth. Zero$\rightarrow$held-out and zero$\rightarrow$counterfactual apply the zero-aligned score to a different ground truth; zero$\rightarrow$zero and the remaining rows match attribution and intervention semantics. Spearman is the mean full-vector correlation across models 000--009 for each behaviour; AUROC (ground-truth top 20) and precision@20 are means over the 20 model--behaviour pairs.}
\label{tab:attr}
\small
\setlength{\tabcolsep}{4.5pt}
\begin{tabular}{llrrrr}
\toprule
 & & \multicolumn{2}{c}{Spearman} & & \\
method & score semantic $\rightarrow$ ground truth & B0 & B1 & AUROC & P@20 \\
\midrule
IG & zero$\rightarrow$zero & 0.777 & 0.614 & 0.936 & 0.473 \\
IG & zero$\rightarrow$held-out & 0.122 & 0.007 & 0.918 & 0.440 \\
IG & held-out$\rightarrow$held-out & 0.565 & 0.385 & 0.947 & 0.505 \\
IG & zero$\rightarrow$counterfactual & 0.218 & $-0.061$ & 0.927 & 0.453 \\
IG & counterfactual$\rightarrow$counterfactual & 0.638 & 0.118 & 0.948 & 0.490 \\
\midrule
first-order EAP & zero$\rightarrow$zero & 0.900 & 0.886 & 0.927 & 0.500 \\
first-order EAP & zero$\rightarrow$held-out & 0.190 & 0.062 & 0.933 & 0.485 \\
first-order EAP & held-out$\rightarrow$held-out & 0.884 & 0.860 & 0.970 & 0.655 \\
first-order EAP & zero$\rightarrow$counterfactual & 0.279 & $-0.033$ & 0.943 & 0.480 \\
first-order EAP & counterfactual$\rightarrow$counterfactual & 0.813 & 0.762 & 0.961 & 0.570 \\
\bottomrule
\end{tabular}
\end{table}

\begin{figure}[t]
\centering
\includegraphics[width=\linewidth]{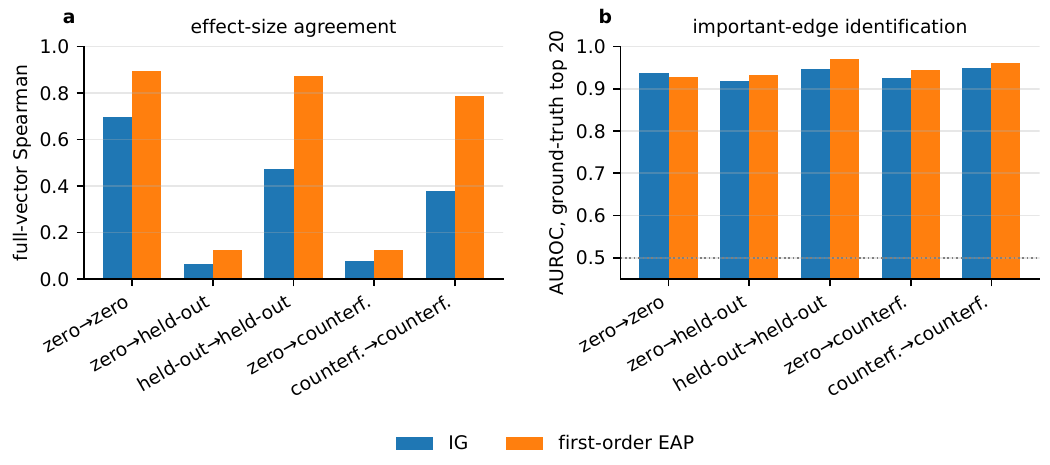}
\caption{Attribution agreement averaged over the 20 model--behaviour pairs. (a) Full-vector Spearman and (b) AUROC for the ground-truth top 20, for zero-aligned and intervention-matched IG and first-order EAP. The zero$\rightarrow$zero bars show the matched zero-ablation case for reference.}
\label{fig:attr}
\end{figure}

Matching the attribution path to the intervention improves continuous-effect agreement, especially for first-order EAP (\cref{tab:attr}; \cref{fig:attr}). For held-out mean replacement, matched EAP reaches Spearman 0.884 for B0 and 0.860 for B1, whereas zero-aligned EAP reaches 0.190 and 0.062; averaged over the 20 model--behaviour pairs this is 0.872 versus 0.126, and the Spearman restricted to the union of ground-truth top-100 connections rises from 0.132 to 0.783. Matched IG reaches 0.565 and 0.385. Under counterfactual replacement, matched EAP reaches 0.813 and 0.762, while matched IG reaches 0.638 for B0 but only 0.118 for B1. Identification of the important connections is much less sensitive to the mismatch: AUROC stays above 0.91 in every row, although precision@20 ranges only from 0.44 to 0.66, so even matched EAP places about 13 of the 20 ground-truth connections in its own top 20. Attribution methods therefore rank the important connections highly under any of these baselines, but their effect sizes are only meaningful for the intervention they were constructed for.

Under held-out-matched attribution, IG and EAP miss 127 and 72 ground-truth top-20 connections outside their own top 40, of which 44.1\% and 50.0\% are inhibitory. The held-out ground-truth top-20 pools are themselves 43--46\% inhibitory, so the misses are not detectably enriched for inhibition. IG completeness is not exact: the median absolute clean-normalized residual is 0.00555 for held-out mean, and near-zero B1 outputs make some normalized residuals ill-conditioned.

\subsection{Single-connection rankings do not give a sufficient circuit}

Exhaustive single-connection interventions give single-component effects, not a description of how components act together. To test whether the top-ranked connections reproduce the behaviour, we score a connection set by the number $\Fc$ of eight clips (standard gratings and their backward-drifting counterparts in four directions) on which the model with all other connections ablated reproduces the full model's decoded direction (within $45^\circ$), magnitude (ratio in $[0.5,2]$) and bounded activity. Under zero ablation of the complement, no greedy set reaches $\Fc\ge6$ within 60 connections in any of the ten models, and top-$k$ sets need a median of 320 connections when ranked by IG or by the true single-connection effect and all 604 when ranked by EAP. Under same-stimulus mean replacement, greedy search reaches $\Fc\ge6$ with a median of 4.5 connections in 16 of 20 cases, but only 3 of those sets contain any directed path from a photoreceptor to a decoder input: the small sets are produced by stimulus leakage, not by a compact circuit (\cref{app:sufficiency}).

% ===========================================================================
\section{Cell-type circuits}
\label{sec:types}

The connection-level results show that attribution depends on intervention semantics, but they define a circuit in fine-grained type-to-type connection classes. We now ask how the mechanistic picture changes when the intervention unit is a biological cell type and when the analysis is extended across the released ensemble. The connection-level experiments and single-type diagnostics use models 000--009. Joint T4/T5 silencing and the tuning-based anatomical, random and greedy cell-type analyses use all 50 released models with the same stimuli, thresholds and search rules.

\paragraph{T4/T5 output silencing.}
Silencing the output of T4a--d and T5a--d reduces the eight-clip score to $\Fc\le2$ in 40 of 50 models (80.0\%; Wilson 95\% CI 67.0--88.8\%). The scores are 0 in 18 models, 1 in 6, 2 in 16, 3 in 2, 4 in 6 and 6 in 2; only the last two still meet the $\Fc\ge6$ reproduction criterion, and eight further models retain the behaviour on three or four clips. This intervention also removes eight of the decoder's 34 input cell types, so it cannot separate dependence on the recurrent circuit from loss of direct decoder input; we report heterogeneous dependence on T4/T5 output under this model intervention, not biological necessity. In single-type silencing on models 000--009, Mi4 is among the ten most influential types in all ten models and Mi1 in nine; Tm1, Tm9, Mi9, Tm3 and CT1 recur, together with photoreceptors, lamina neurons and several types outside the canonical pathway (TmY15, T2, C2). Between 2 and 6 of each model's ten most influential types belong to the anatomical ON or OFF motion pathway \citep{borst2020}. IG with respect to per-type gains recovers the single-type ranking with mean AUROC 0.80 (Spearman 0.60), lower than at the connection level. Keeping only a small set of cell types cannot be evaluated at the decoder, because silencing most of its 34 inputs changes the readout itself (\cref{app:sufficiency}).

\paragraph{A model-relative T4/T5 tuning criterion.}
To remove the decoder from the test, we measure the central T4a--d and T5a--d cells with gratings at eight directions (4\,Hz, contrast 1). The response of a cell is its signed mean activity over the analysis window. We estimate the preferred direction from the response-weighted vector over the eight gratings, take the sampled gratings nearest that angle and its opposite, and compute
\begin{equation}
\mathrm{DSI} = \frac{R_{\text{pref}} - R_{\text{null}}}{|R_{\text{pref}}| + |R_{\text{null}}|}.
\end{equation}
A subtype passes if its DSI is at least half the full model's and its preferred direction is within $45^\circ$ of the full model's. The tuning score $\TF\in\{0,\dots,8\}$ counts passing subtypes. Three properties limit what this score means. First, it is relative to each model's own tuning, not to biology. Against the canonical subtype layout (a=$180^\circ$, b=$0^\circ$, c=$90^\circ$, d=$270^\circ$ for T4 and T5; \citealp{maisak2013}), 190 of 400 preferred directions lie within $45^\circ$ (T4: 109/200; T5: 81/200). This is not specific to our grating measure or angle convention: with flyvis's own moving-edge stimuli and preferred-direction routine (ON edges for T4, OFF edges for T5), agreement is again 190/400 (T4: 115/200; T5: 75/200), and none of the seven other rotations and reflections of the canonical layout does better under either measure (\cref{app:tuning}). Under the edge measure, agreement is higher in models with lower legacy packaged task-error values (Spearman $-0.33$ between that release scalar and the number of aligned subtypes, $p=0.02$; 51/80 in models 000--009). The scalar's exact generator and held-out status are not retained, so this association is descriptive and is not itself a verified performance comparison; \citet{lappalainen2024} separately reported that better task performance accompanies more realistic tuning. Second, the absolute-value denominator saturates: DSI equals 1 in 140/400 full-model cases, including all 38 cases in models 000--009 in which preferred and null responses have opposite signs. Third, the criterion is one-sided: it places no upper bound on DSI, response amplitude or overall activity. We therefore describe $\TF\ge6$ as meeting a model-relative tuning criterion, not as physiological sufficiency.

\paragraph{Ensemble results.}
With only the anatomical motion pathways active (32 types including photoreceptors, lamina and T4/T5; \cref{app:details}), $\TF\ge6$ in 4 of 50 models (\cref{tab:types}); the score is 0 in 14 models, 1--3 in 31 and 5 or 6 in 5. Size-matched random sets, which keep the 21 photoreceptor, lamina and T4/T5 types and draw 11 of the 32 types outside the anatomical set, score 0 in 989 of 1{,}000 draws and 1 in the other 11, so the anatomical set exceeds all 20 of its random draws in 35 of 50 models mostly by scoring at all. Greedy selection from the 21-type core reaches $\TF\ge6$ within 15 additions in 25 of 50 models, after a median of 7 additions (range 3--15). Mi1 appears in all 25 successful sets, TmY15 in 23, Mi4 in 17 and Mi9 in 16, and at least one CT1 compartment in 20; the remaining composition varies across models.

\begin{table}[t]
\centering
\caption{Cell-type results over all 50 released models, with Wilson 95\% intervals.}
\label{tab:types}
\small
\setlength{\tabcolsep}{5pt}
\begin{tabular}{p{0.57\linewidth}cc}
\toprule
outcome & models & rate (95\% CI) \\
\midrule
T4/T5 jointly silenced, $\Fc\le2$ & 40/50 & 80.0\% (67.0--88.8\%) \\
T4/T5 jointly silenced, $\Fc\ge6$ & 2/50 & 4.0\% (1.1--13.5\%) \\
anatomical set, $\TF\ge6$ & 4/50 & 8.0\% (3.2--18.8\%) \\
anatomy exceeds all 20 matched random draws & 35/50 & 70.0\% (56.2--80.9\%) \\
greedy set reaches $\TF\ge6$ within 15 additions & 25/50 & 50.0\% (36.6--63.4\%) \\
\bottomrule
\end{tabular}
\end{table}

\paragraph{Matched cell-type and connection-class search.}
The decoder-level connection search and the tuning-level cell-type search use different targets, so they cannot show an effect of the intervention unit. We therefore reran models 000--009 with the same T4/T5 tuning target, tie break and timing at both levels. The connection representation that matches cell-type output gains is the set of connections whose \emph{source} type is active: this reproduces the cell-type traces to relative error at most $2.65\times10^{-7}$, and the 21-type core corresponds to a common core of 184 connections.

Cell-type search reaches $\TF\ge6$ in 6 of 10 models, after 4, 4, 9, 6, 9 and 9 additions. A forward search over single connections reaches the criterion in 0 of 10 models within 60 additions, but its integer objective plateaus and its budget is smaller than the 71--164 connections that the successful type sets add, so this does not show that no smaller connection set exists (\cref{app:sufficiency}). We instead expand each successful type set into its source-active connections and greedily prune connections outside the common core. Pruning removes 57--149 of the 71--164 added connections (74--91\%) and leaves 13--31, or 197--215 in total (\cref{tab:pruned}). All six endpoints retain $\TF\ge6$ in three independent evaluations and are one-step irreducible. Three caveats apply. A type intervention gates many connections at once and the type sets were not themselves pruned, so the comparison of 4--9 types with 13--31 connections is partly structural; the informative result is how much of each type-derived connection set can be discarded. The greedy path is not unique: 455 of 615 evaluated pruning steps were numerically ambiguous at the frozen scale; these 615 evaluations comprise 609 accepted removals plus one terminating rejected step in each of the six models. The endpoints are greedy, one-step irreducible sets rather than globally minimal circuits. Finally, for each subtype in the six first-success \emph{cell-type} sets, a post-hoc sensitivity check additionally required
$0.5\le (R_{\text{pref}}-R_{\text{null}})_{\text{set}}/(R_{\text{pref}}-R_{\text{null}})_{\text{full}}\le2$ relative to that subtype's full-model value. The augmented scores are 5, 6, 5, 7, 4 and 1 for models 001, 003, 004, 005, 007 and 009, respectively; only models 003 and 005 remain at $\TF\ge6$. This check was not performed independently on the pruned connection endpoints.

\begin{table}[t]
\centering
\caption{Matched T4/T5-criterion results for the six models in which the cell-type search reaches $\TF\ge6$. ``Source types retained'' counts added source types with at least one outgoing connection left after pruning.}
\label{tab:pruned}
\small
\setlength{\tabcolsep}{3.8pt}
\begin{tabular}{lrrrrr}
\toprule
model & added types & source types retained & added connections & after pruning & total connections \\
\midrule
001 & 4 & 4 & 101 & 13 & 197 \\
003 & 4 & 4 & 71  & 14 & 198 \\
004 & 9 & 9 & 153 & 26 & 210 \\
005 & 6 & 6 & 120 & 18 & 202 \\
007 & 9 & 9 & 117 & 31 & 215 \\
009 & 9 & 6 & 164 & 15 & 199 \\
\bottomrule
\end{tabular}
\end{table}

% ===========================================================================
\section{Discussion}
\label{sec:discussion}

Across the paper, the same reliability issue appears at three levels. External behaviour is not uniquely family-specific in the tested comparisons; attribution effect sizes depend on the intervention used to define the ground truth; and criterion-preserving circuits depend on the representation and trained ensemble member. These observations narrow what the connectome constraint alone can support while preserving its main practical advantage: mechanistic hypotheses can be stated on named biological components and tested by intervention.

\paragraph{Model-family comparison.}
The CNN controls share the task, Sintel rendering and target, loss and input lattice with flyvis but not a verified training split; they also differ in recurrence, learned time constants, parameter sharing, decoder, temporal context and training schedule, so differences between the families cannot be assigned to connectome wiring. With that caveat, the families are more alike than different on our tests: both reverse under reverse-phi in selected bands, neither responds to stationary sawtooth gradients like the fly, and the adversarial margin of flyvis exceeds that of a parameter-matched CNN but not clearly that of larger CNNs. A practical advantage of the connectome-constrained representation is experimental access: named cell types and connections can be intervened on and compared with biological perturbations.

\paragraph{Interventions define the question.}
Zero-aligned IG and EAP identify important connections under every replacement we tested, but their effect sizes agree with the ground truth only when the attribution is constructed for the same intervention. The discrepancy is therefore largely a mismatch between the causal question and the attribution construction rather than a failure of the methods, in line with the view that the ablation baseline defines the question being asked \citep{li2024optimal,chan2022causal}. Likewise, how compact a circuit looks depends on the unit of intervention and the search procedure: the cell-type and connection-class results describe how criterion-preserving structure is distributed across coarse and fine units, not an equal-unit compression comparison.

\paragraph{Degeneracy across the ensemble.}
The greedy cell-type search meets the tuning criterion in 25 of 50 models and fails in the other 25, and T4/T5 silencing does not reduce the score to $\Fc\le2$ in ten models. We detect no association between either split and the release package's legacy task-error ranking or clean-grating accuracy. For the legacy scalar, rank-biserial effects for the two splits are $-0.060$ and $-0.168$ (two-sided Mann--Whitney $p=.780$ and $.313$); for clean-grating accuracy they are $-0.063$ and $0.005$ ($p=.767$ and $.984$), with identical group medians of clean-grating accuracy (0.9375). We use the packaged scalar only as a within-release ranking variable because its generator and held-out provenance are not retained (\cref{app:validation}); the nonsignificant comparisons do not establish equivalent task quality. Agreement with canonical T4/T5 preferred directions, in contrast, has a weak descriptive association with the legacy scalar (\cref{sec:types}). This variation is consistent with retraining evidence that task performance does not uniquely identify biologically plausible flyvis solutions \citep{karaneen2026}, with underdetermination theory \citep{beiran2025}, and with explanation non-identifiability \citep{meloux2025} and biological degeneracy \citep{prinz2004,marder2006}. Mechanistic claims drawn from a single trained model are claims about one member of an ensemble of trained solutions; stimuli designed to separate these solutions \citep{golan2020}, evaluated against recordings, are a natural next step.

\paragraph{Limitations.}
All tests use one architecture and its released ensemble. The released models' launch-time training split and the exact generator of their packaged task-error scalar are not recoverable. Consequently, the common original-16 evaluation in \cref{tab:models} is not a verified held-out test for flyvis and cannot establish a fair shared held-out generalization comparison with the CNNs. The CNNs were trained for 30{,}000 steps against flyvis's 250{,}000. Most stimuli are periodic gratings; attacks use one direction and frequency per condition, a single random start and the central-column readout, and low-contrast CNN cells rest on one or two seeds. Sawtooth luminances are normalized rather than calibrated to the fly experiments, and no eye movements were simulated. Held-out means exclude target movies but reuse the same stimulus families and three calibration directions. Attribution is evaluated in the ten models ranked best by the legacy packaged scalar and two readouts, one of which (B1) is small in most models; first-order EAP is a local approximation. CUDA nondeterminism makes near-zero ranks and some greedy choices non-unique. Output silencing is a model intervention, not cell removal, and lesion-style effects should not be equated with a complete mechanism \citep{jonas2017}. The matched-random comparison uses 20 draws per model. The remaining caveats on the reverse-phi stimulus, the tuning criterion and the pruned connection sets are stated with the corresponding results.

\section{Conclusion}
\label{sec:conclusion}

Connectome constraints make mechanistic hypotheses directly testable, but in flyvis the answers are conditional. The tested model-family behaviours do not isolate an effect of wiring; attribution effect sizes depend on the intervention they are matched to; and criterion-preserving circuits vary with representation and ensemble member. The value of the connectome constraint is therefore not that it makes a mechanism unique, but that it makes competing mechanistic claims concrete enough to intervene on and compare. Such claims should state the model family, intervention, circuit representation and trained solution for which they hold.

\subsubsection*{Code and data availability}
All code, stimulus definitions, job scripts, logs and intermediate result files will be released publicly upon publication.

\ifnamedversion
\subsubsection*{Acknowledgments}
The author thanks Daisy Zhe Wang for advising and Dzmitry Y. Kasinets for discussions. The author acknowledges University of Florida Research Computing and HiPerGator for computational resources supporting this work.
\fi

\bibliography{references}
\bibliographystyle{tmlr}

\appendix
\crefalias{section}{appendix}
\crefalias{subsection}{subappendix}
\section{Predictions and analysis gates recorded before the runs}
\label{app:prereg}

\Cref{tab:prereg} lists directional predictions and analysis gates that were written into the experiment scripts before the corresponding runs. They were not externally timestamped preregistrations. No outcome prediction was recorded before the three-level reverse-phi evaluation, which was introduced as a stimulus-validity control.

\begin{table}[h]
\centering
\caption{Directional predictions and analysis gates recorded in scripts before execution. Results over all 50 models use the same criteria as the ten-model results. Three entries (population-vector readout, timestep check, anatomical set without lamina) are reported only here; predictions about analyses outside this paper are omitted.}
\label{tab:prereg}
\small
\begin{tabular}{p{0.44\linewidth}p{0.28\linewidth}p{0.20\linewidth}}
\toprule
prediction or gate & outcome & status \\
\midrule
Population-vector readout from tabulated T4/T5 preferred directions: $\ge35/50$ models reverse (quarter-period stimulus) & 2/50; readout near chance on standard gratings (26--30\% against 25\%) & failed; readout not used \\
Quarter-period two-level reverse-phi: $\ge35/50$ reverse & 49/50; phase-equivalent to same-speed backward motion & not diagnostic (\cref{app:artefacts}) \\
Two-level contrast-inverting grating, 17-pixel step: $\ge35/50$ reverse at one of the three slowest velocities & 42--48/50 for the $30^\circ$ period & met; phase-equivalent to translation (\cref{app:artefacts}) \\
Timestep $1/50$\,s changes no decoder verdict & 0/10 changed & met \\
IG/EAP AUROC $\le0.75$ on B1 in $\ge6/10$ models & mean AUROC 0.94--0.95 & failed \\
Misses enriched for inhibitory connections & held-out-matched misses 44.1\%/50.0\%; held-out top-20 pools about 43--46\% inhibitory & not supported \\
Misses enriched for recurrent connections & uninformative (97.7\% recurrent) & not evaluable \\
T4/T5 silencing gives $\Fc\le2$ in all models & 9/10 in models 000--009; 40/50 overall & failed \\
Four-level static pattern: $\ge35/50$ flyvis models, CNNs not & max.\ 14/50; CNNs max.\ 2/6 & failed / met \\
Sawtooth gradient: $\ge35/50$ flyvis models, sign flips with polarity & 3/50 per polarity, 0/50 both & failed \\
Anatomical set gives $\TF\ge6$ in $\ge7/10$ models & 1/10 in models 000--009; 4/50 overall & failed \\
Anatomical set without lamina gives $\TF\le2$ in all & 8/10 & failed \\
Random type sets: median $\TF\le3$ & 0 & met \\
Anatomical set exceeds size-matched random draws & 8/10 in models 000--009; 35/50 exceed all 20 draws overall & not universal \\
Greedy type set reaches $\TF\ge6$ within 10 added in $\ge7/10$ & 6/10 in models 000--009; 25/50 within 15 overall & failed \\
Greedy connection set reaches $\Fc\ge6$ within 30 additions & 0/10 within 60 additions & failed \\
\bottomrule
\end{tabular}
\end{table}

\section{Evaluation artefacts}
\label{app:artefacts}

\paragraph{Periodic phase equivalence in reverse-phi.}
For an ideal 50\%-duty binary mask $M$, contrast complementation satisfies $1-M(x)=M(x+\lambda/2)$. A two-level grating whose contrast inverts at each spatial step therefore also has an ordinary shifted-grating description with an aliased displacement, and is a motion-correspondence control rather than a clean ON/OFF-interaction stimulus. Exhaustive integer translations over one spatial period recover exact half-period equality for all 209 polarity-flipped frames at the even 388-pixel ($90^\circ$) period; the odd 129-pixel ($30^\circ$) period leaves a one-pixel discretization defect (minimum MSE 0.00775). The three-level stimulus of \cref{sec:illusions} is $I(x,t)=0.5+0.5\,s(t)\,M(x-\phi(t))$, where only the non-gray stripes change sign. It is not pixelwise equal to a translated matched-control frame (0/836; minimum MSE 0.25), but after subtracting each frame's spatial mean it equals the two-level moving pattern at half contrast; the remaining difference is a spatially uniform luminance flicker. Unlike the two-level construction, polarity-rectified bright-only or dark-only views of the three-level stimulus signal the nominal direction.

\Cref{tab:revphi_oldnew} compares reversal counts for the two constructions under the same models, periods, velocities, decoder and criterion. The comparison conflates the polarity structure with the halved moving-pattern contrast and the added flicker, so the change in counts cannot be assigned to any one of them. The quarter-period two-level clips used in the eight-clip score $\Fc$ are phase-equivalent to same-speed backward drift and are interpreted only as backward-drifting gratings.

\begin{table}[h]
\centering
\caption{Reversal counts for the two-level contrast-inverting grating $\rightarrow$ the three-level reverse-phi stimulus, by velocity ($^\circ$/s). CNN denominators are five checkpoints per architecture.}
\label{tab:revphi_oldnew}
\small
\setlength{\tabcolsep}{4pt}
\begin{tabular}{llrrrrrr}
\toprule
family & period & 197.2 & 98.6 & 65.7 & 39.4 & 19.7 & 7.89 \\
\midrule
flyvis (50) & $30^\circ$ & 34$\to$14 & 48$\to$46 & 48$\to$25 & 48$\to$32 & 48$\to$37 & 42$\to$20 \\
flyvis (50) & $90^\circ$ & 7$\to$3 & 48$\to$32 & 48$\to$37 & 39$\to$19 & 26$\to$1 & 18$\to$3 \\
CNN small (5) & $30^\circ$ & 2$\to$0 & 4$\to$0 & 5$\to$2 & 4$\to$3 & 4$\to$1 & 1$\to$1 \\
CNN small (5) & $90^\circ$ & 0$\to$0 & 5$\to$4 & 5$\to$5 & 5$\to$4 & 3$\to$0 & 0$\to$0 \\
CNN medium (5) & $30^\circ$ & 0$\to$0 & 5$\to$2 & 5$\to$3 & 5$\to$5 & 5$\to$3 & 2$\to$3 \\
CNN medium (5) & $90^\circ$ & 0$\to$0 & 5$\to$5 & 5$\to$5 & 5$\to$0 & 5$\to$0 & 5$\to$2 \\
CNN large (5) & $30^\circ$ & 4$\to$0 & 5$\to$2 & 5$\to$2 & 5$\to$2 & 5$\to$2 & 3$\to$2 \\
CNN large (5) & $90^\circ$ & 0$\to$0 & 5$\to$4 & 5$\to$3 & 4$\to$0 & 5$\to$0 & 4$\to$0 \\
\bottomrule
\end{tabular}
\end{table}

\paragraph{One-sided retained-output metric.}
A retained-output diagnostic scored a set $S$ by $M(S)/M_{\text{clean}}$ on one stimulus and counted the set as successful when this ratio exceeded 0.8. Greedy search under this metric found sets of 2 to 8 connections. For the resulting high-ranked cores, however, the ratio ranged from below 0 to above 3, and in some curves it exceeded $10^3$: removing most of the network can amplify or reverse a small normalized output. Under the eight-clip score $\Fc$, the same sets score 0 to 2 of 8. This motivates two-sided magnitude, direction and activity constraints when a metric is meant to establish output-level faithfulness. The T4/T5 tuning score of \cref{sec:types} answers a narrower question and is itself one-sided in amplitude; its sensitivity to a two-sided band is reported there.

\paragraph{Mean-current leakage.}
Same-stimulus mean-current ablation injects, into every ablated connection, its time-averaged current from the clean run on the target clip. These currents differ between clips and carry information about the stimulus. Sets found under this ablation reproduce the full model on most clips while, in 13 of 16 cases, containing no directed path from photoreceptors to the decoder. Under this same-stimulus ground truth, zero-aligned IG and EAP miss 166 and 149 ground-truth top-20 connections outside their top 40, of which 56.0\% and 63.1\% are inhibitory; the apparent inhibitory enrichment disappears under held-out-matched attribution (\cref{sec:circuits}). We therefore use same-stimulus mean only as a leakage diagnostic and currents averaged over held-out calibration movies as the primary mean replacement.

\section{Stimulus and implementation details}
\label{app:details}

\paragraph{Gratings.}
Gratings are rendered on flyvis's canvas at 25 pixels per ommatidium ($5.8^\circ$ per ommatidium). Adversarial, B0 and tuning experiments use square-wave gratings of period 152 pixels (6.08 ommatidia) advancing by a quarter period per step. The reverse-phi experiment uses three luminance levels: alternating stripes remain at gray 0.5 while the non-gray stripes are 0 or 1 and switch polarity after each 17-pixel ($3.944^\circ$) update. Periods are 129 and 388 pixels ($29.928^\circ$, $90.016^\circ$), with update intervals of 1, 2, 3, 5, 10 and 25 frames at 50\,Hz (197.2, 98.6, 65.73, 39.44, 19.72 and 7.888$^\circ$/s). Matched controls have identical geometry, background, starting polarity, displacement, timing, duration and sampling. In the reversal criterion, the directions that pass the standard-control test and those that pass the reversal test need not coincide. The held-out calibration movies are standard and two-level contrast-inverting gratings at $90^\circ$, $180^\circ$ and $270^\circ$. The sawtooth stimulus has a $45^\circ$ period, both polarities, 1\,s of mean grey before a 5\,s presentation, and normalized luminance in $[0,1]$; its analysis window runs from 0.5\,s after onset to the end.

\paragraph{CNN training.}
CNNs were trained with Adam and cosine decay for 30{,}000 steps, keeping the best validation checkpoint. A learning criterion fixed in the training script was not met by four of the first six checkpoints at an initial learning rate of $10^{-3}$; as specified in the script, all three sizes were then retrained at $3\times10^{-4}$, and all 15 reported checkpoints come from that setting. flyvis checkpoints were trained by \citet{lappalainen2024} for 250{,}000 Adam steps with a stepwise $5\times10^{-5}$ to $5\times10^{-6}$ schedule.

\paragraph{Common evaluation and legacy release scalar.}
\label{app:validation}
\Cref{tab:models} reports a common evaluation of flyvis models 000--009 and all CNNs on the original 16-subsequence Sintel evaluation set with augmentation disabled. Each input uses the first 19 original frames at 24\,Hz: luminance is converted to gray, divided by 255, spatially mean-sampled onto the hexagonal lattice and piecewise-constant resampled to 50\,Hz. Target flow is divided by image height, has its vertical sign inverted, is spatially summed by the eye kernel and linearly interpolated to 50\,Hz. This produces 40 evaluated frames. For flyvis, a constant-gray recurrent warm-up of 0.25\,s is recomputed for each subsequence; the feedforward CNNs have no recurrent warm-up. With batch size one, the per-subsequence $\ell_2$ scalar is $\sqrt{\sum_{t,c,h}(\hat y-y)^2}$ over 40 frames, two flow components and 721 hexals, and the table gives the equal arithmetic mean of the 16 scalars.

The original evaluation scenes are disjoint from the documented CNN training split. The released flyvis package does not retain its launch-time training split. A historical random fold is the best-supported reconstruction, but its training side overlaps 11 of the original 16 subsequences and all six original-evaluation scenes, so the flyvis values are not demonstrably held out. Table~\ref{tab:models} is therefore descriptive across model families rather than a fair shared held-out generalization comparison.

The release package separately stores a scalar near 5.3 under \texttt{validation/loss} and \texttt{validation\_loss}. Under a frozen common evaluator this scalar is close to EPE on the original evaluation set and is incompatible with current $\ell_2$ norm or EPE on reconstructed random fold~1. The package metadata was manually written, and the scalar's exact generator, configuration, checkpoint-selection history and training-held-out status were not retained. We therefore use it only as a legacy within-release ranking variable; we do not call it a verified held-out validation loss or claim to reproduce its exact generator.

\paragraph{Cell indices.}
Tuning was measured at the central-column cells of T4a--d and T5a--d (flyvis node indices 24399, 25120, 25841, 26562, 27283, 28004, 28725, 29446).

\paragraph{Anatomical pathway set.}
R1--R8, L1--L5, Mi1, Tm3, Mi4, Mi9, C3, CT1(M10), CT1(Lo1), Tm1, Tm2, Tm4, Tm9, T4a--d and T5a--d, with Am intact.

\paragraph{Eight-clip score.}
Standard gratings at 4\,Hz and quarter-period contrast-inverting gratings at 3.125\,Hz (phase-equivalent to same-speed backward drift), each at $0^\circ$, $90^\circ$, $180^\circ$ and $270^\circ$, contrast 1, 50\,Hz. A clip counts as reproduced when the decoded direction lies within $45^\circ$ of the full model's, the magnitude ratio lies in $[0.5,2]$, and mean absolute activity stays below five times that of the full model.

\paragraph{Compute.}
GPU jobs used one NVIDIA B200 or L4 and five CPU cores.

\section{Attack transfer and input-gradient similarity}
\label{app:transfer}

\begin{figure}[h]
\centering
\includegraphics[width=\linewidth]{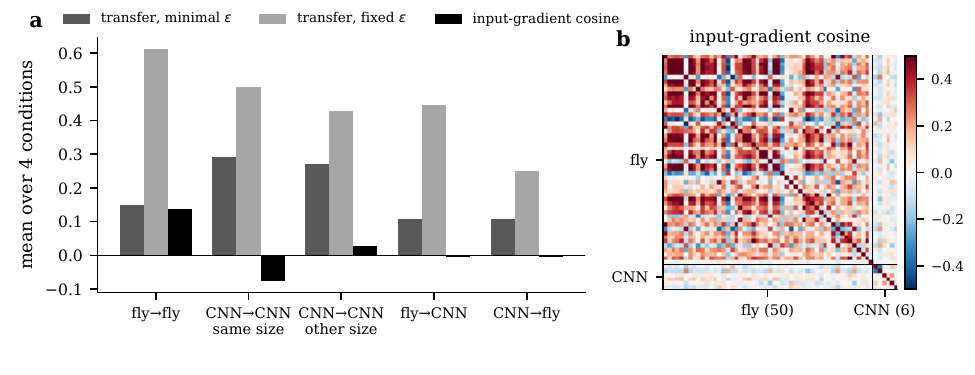}
\caption{(a) Transfer and input-gradient similarity by source and target family. (b) Input-gradient cosine similarity between all 56 models, averaged over conditions; rows and columns are ordered flyvis 000--049, then CNNs.}
\label{fig:transfer}
\end{figure}

We computed full $56\times56$ transfer and gradient-similarity matrices over all 50 flyvis models and six CNNs (the first two seeds of each size) at contrasts 0.25 and 0.1 and directions $0^\circ$ and $90^\circ$ (\cref{tab:transfer}, \cref{fig:transfer}). Perturbations scaled to each source's own $\epsilon^\ast$ transfer less often between flyvis models (0.15) than between CNN seeds of the same size (0.29). At a fixed bound (0.08 at contrast 0.25, 0.032 at contrast 0.1), flyvis-to-flyvis transfer is 0.61 and same-size CNN transfer 0.50. The same-size CNN comparison contains only six ordered pairs per condition (24 binary outcomes across the four conditions), and the within-family ordering changes with perturbation normalization, so we treat these rates descriptively. Hierarchical clustering of any of these matrices does not separate the families. Input gradients are more similar among flyvis models (mean cosine 0.14) than among CNN seeds of the same size ($-0.08$), consistent with the flyvis models sharing one architecture and training objective.

\begin{table}[h]
\centering
\caption{Transfer rates and input-gradient cosine similarity, averaged over four conditions.}
\label{tab:transfer}
\small
\begin{tabular}{lccccc}
\toprule
 & fly$\to$fly & CNN$\to$CNN (same size) & CNN$\to$CNN (other size) & fly$\to$CNN & CNN$\to$fly \\
\midrule
transfer, minimal $\epsilon$ & 0.148 & 0.292 & 0.271 & 0.109 & 0.107 \\
transfer, fixed $\epsilon$   & 0.612 & 0.500 & 0.427 & 0.445 & 0.249 \\
gradient cosine              & 0.137 & $-0.076$ & 0.027 & $-0.005$ & $-0.005$ \\
\bottomrule
\end{tabular}
\end{table}

\section{Stationary patterns}
\label{app:static}

\Cref{tab:static} gives the full counts behind the stationary-pattern results summarized in \cref{sec:illusions}.

\begin{table}[h]
\centering
\caption{Stationary patterns. Top: sawtooth gradients \citep{agrochao2020}, number of models meeting the direction and magnitude criterion (left-to-right / right-to-left polarity) and median response relative to a moving grating. Bottom: four-level peripheral-drift sequence (black, dark grey, white, light grey; \citealp{conway2005}), number of models reporting the ordering-predicted direction (forward / reverse ordering) in the sustained window, by spatial period in ommatidia.}
\label{tab:static}
\small
\begin{tabular}{lccc}
\toprule
\multicolumn{4}{l}{\emph{Sawtooth gradient, $45^\circ$ period, 5\,s}} \\
family & criterion met & met for both polarities & median response ratio \\
\midrule
flyvis (50)     & 3 / 3 & 0 & 0.023 / 0.026 \\
CNN small (5)   & 0 / 0 & 0 & 0.014 / 0.019 \\
CNN medium (5)  & 0 / 0 & 0 & 0.022 / 0.016 \\
CNN large (5)   & 0 / 0 & 0 & 0.055 / 0.031 \\
\midrule
\multicolumn{4}{l}{\emph{Four-level peripheral-drift sequence}} \\
family, presentation & period 4 & period 6 & period 8 \\
\midrule
flyvis (50), held    & 3 / 1 & 4 / 0 & 4 / 1 \\
flyvis (50), flashed & 1 / 4 & 5 / 4 & 14 / 2 \\
CNN (6), held        & 0 / 0 & 0 / 0 & 0 / 0 \\
CNN (6), flashed     & 1 / 0 & 0 / 1 & 0 / 2 \\
\bottomrule
\end{tabular}
\end{table}

\section{Sufficiency diagnostics}
\label{app:sufficiency}

\begin{figure}[h]
\centering
\includegraphics[width=\linewidth]{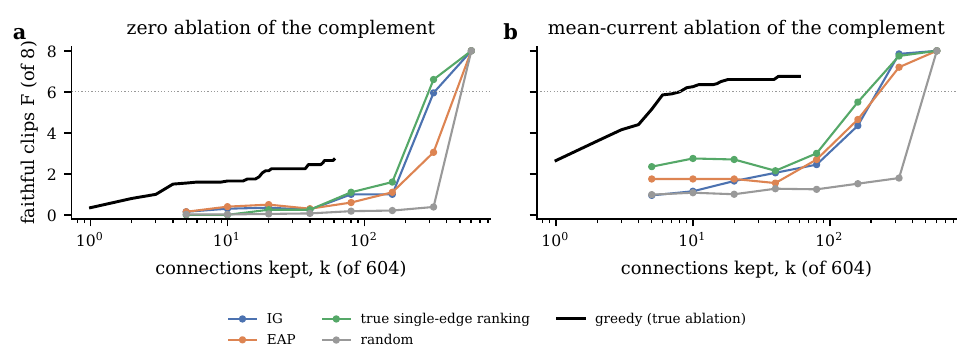}
\caption{Decoder-level connection sufficiency. Mean faithful clips $\Fc$ (of 8) when only $k$ connections are kept, over 10 models and two ranking stimuli. (a) Complement zero-ablated. (b) Complement replaced by same-stimulus clean mean currents; the apparently small sets are leakage-prone and mostly lack an input-to-output path. Legend: ``true single-edge ranking'' is the ranking by true single-connection effect; ``greedy (true ablation)'' is greedy forward selection evaluated by direct intervention.}
\label{fig:faith}
\end{figure}

\paragraph{Connection-level sufficiency at the decoder.}
We compare top-$k$ connection sets ranked by IG, EAP and the true single-connection effect with random sets and with greedy forward selection that maximizes the eight-clip score $\Fc$ directly, the latter over the union of the 100 connections with the largest zero and the largest same-stimulus-mean effects. Under zero ablation of the complement, no greedy set reaches $\Fc\ge6$ within 60 connections in any of the ten models (\cref{fig:faith}a). Top-$k$ sets need a median of 320 connections for IG (range 80--604), 604 for EAP (320--604) and 320 for the true single-connection ranking (160--604). Under same-stimulus mean replacement, greedy search reaches $\Fc\ge6$ with a median of 4.5 connections in 16 of 20 cases (\cref{fig:faith}b), but only 3 of these sets contain any directed path from a photoreceptor to a decoder input.

\paragraph{Cell-type sufficiency at the decoder.}
In models 000--009, keeping only a greedily selected set of up to 20 cell types in addition to the photoreceptors gives $\Fc\le4$ in every model, and the anatomical ON and OFF pathways alone give $\Fc$ of 0 to 5. Because the decoder reads 34 cell types, silencing most of them also changes the decoder's input distribution, so this test conflates the circuit's computation with the readout. Under the tuning criterion, the decoder's 34 input types and the decoder-optimized type sets score $\TF=0$.

\begin{figure}[h]
\centering
\includegraphics[width=\linewidth]{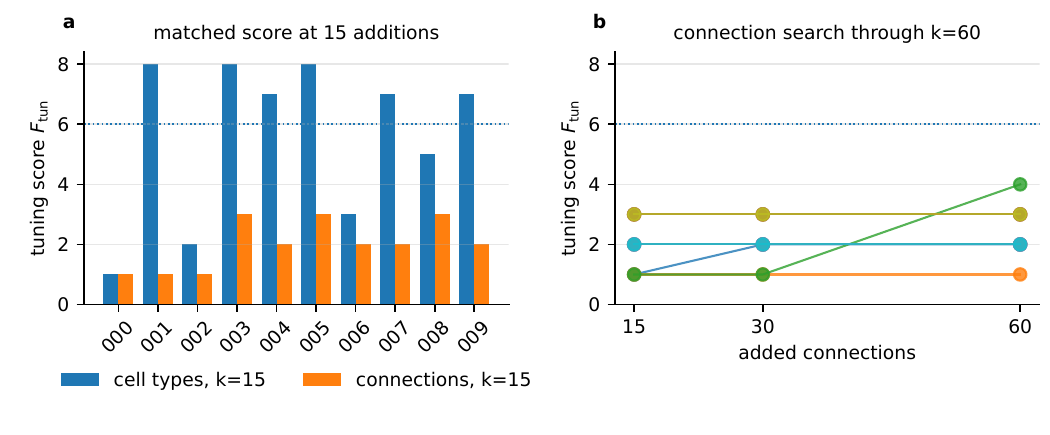}
\caption{Forward search under the matched T4/T5 tuning criterion in models 000--009. (a) Tuning score after 15 additions for cell-type and single-connection searches. (b) Connection-search scores at 15, 30 and 60 additions; each colored trajectory is one model and overlapping trajectories coincide. The horizontal line marks $\TF\ge6$.}
\label{fig:types}
\end{figure}

\paragraph{Forward search under the matched tuning criterion.}
\Cref{fig:types} shows the forward searches of \cref{sec:types}. Enabling all 604 connections reproduces native activity to relative error at most $5.38\times10^{-7}$ and gives $\TF=8$. The single-connection forward search plateaus below the criterion and its 60-connection budget is smaller than several type-derived connection sets, so it is a search diagnostic and not a lower bound on connection-set size.

\section{Canonical preferred directions under two response measures}
\label{app:tuning}

\Cref{tab:frames} counts, over the 400 subtype--model cases in the 50 intact models, how many preferred directions lie within $45^\circ$ of the canonical layout (a=$180^\circ$, b=$0^\circ$, c=$90^\circ$, d=$270^\circ$) and of its seven other rotations and reflections. The grating measure is the one used for $\TF$. The moving-edge measure uses flyvis's own stimuli (12 directions, six speeds, ON edges for T4 and OFF edges for T5) and its preferred-direction routine. A uniformly distributed preferred direction would give 100/400. The identity frame is best under both measures, so the 190/400 agreement is not explained by a rotated or mirrored angle convention; the next-best frame exchanges the a and b subtypes. Per-model counts of aligned subtypes range from zero to eight under both measures; one model aligns all eight under the grating measure and two under the edge measure.

\begin{table}[h]
\centering
\caption{Subtype--model cases (of 400) within $45^\circ$ of the canonical T4/T5 layout after each frame transform. Mirroring is $\theta\to-\theta$, applied before the rotation.}
\label{tab:frames}
\small
\begin{tabular}{lrr}
\toprule
transform of canonical layout & gratings & moving edges \\
\midrule
identity & 190 & 190 \\
rotation $90^\circ$ & 53 & 61 \\
rotation $180^\circ$ & 92 & 101 \\
rotation $270^\circ$ & 65 & 48 \\
mirror & 126 & 122 \\
mirror, rotation $90^\circ$ & 57 & 49 \\
mirror, rotation $180^\circ$ (a$\leftrightarrow$b) & 156 & 169 \\
mirror, rotation $270^\circ$ & 61 & 60 \\
\bottomrule
\end{tabular}
\end{table}

\end{document}